\documentclass[twocolumn,showpacs,preprintnumbers,amsmath,amssymb]{revtex4}

\usepackage{graphicx}
\usepackage{dcolumn}
\usepackage{bm}

\newcommand{\be}{\begin{equation}}
\newcommand{\en}{\end{equation}}
\newcommand{\bea}{\begin{eqnarray}}
\newcommand{\ena}{\end{eqnarray}}

\newcommand{\hbo}{\hbox to 1 true cm {\hfill } }

\begin{document}


\title{ A gauge invariant cluster algorithm for the Ising spin glass } 

\author{K.~Langfeld$^{a,b}$}
\author{M.~Quandt$^a$}
\author{W.~Lutz$^a$}
\author{H.~Reinhardt$^a$ }

\affiliation{%
\bigskip
$^a$Institut f\"ur Theoretische Physik, Universit\"at T\"ubingen \\
Auf der Morgenstelle 14, D-72076 T\"ubingen, Germany. \\
}%
\affiliation{%
$^b$School of Mathematics \& Statistics, 
Plymouth, PL4 8AA, United Kingdom. } 

\date{\today}

\begin{abstract}
The frustrated Ising model in two dimensions is revisited. The frustration
is quantified in terms of the number of non-trivial plaquettes which is
invariant under the Nishimori gauge symmetry. The exact ground state energy is
calculated using Edmond's algorithm. A novel cluster algorithm is designed 
which
treats gauge equivalent spin glasses on equal footing and allows for efficient
simulations near criticality. As a first application, the specific heat
near criticality is investigated.
\end{abstract}

\pacs{ 11.15.Ha, 12.38.Aw, 12.38.Gc }
\keywords{ frustration, spin-glass, critical phenomenon }
\maketitle

Spin glasses~\cite{Binder86} are magnetic materials in which the magnetic moments
are subject to ferromagnetic or anti-ferromagnetic interactions, depending
on the position of the moments within the sample. The system is \emph{frustrated}
in the sense that the arrangement of spins which minimizes the total energy
cannot be determined by considering a \emph{local} set of spins.
Stated differently, the change of a single spin might cause a reordering of many
spins when the system relaxes towards a new minimum of energy~\cite{Alava98}.
Spin glasses undergo a freezing transition to a state where the order is
represented by clusters of spins with mixed orientations. The relaxation times
towards equilibrium are typically very large, which impedes efficient
simulations.

\vskip 2mm
Many efforts have been undertaken to explore equilibrium properties
of spin glasses by means of Monte Carlo
simulations~\cite{Binder79,Binder80,Young82,Ogielski85,Huse85,Rieger96,Matsubara97,Shirakura97}.
Thereby, many insights have been obtained from the most simple case
of the 2d Ising model on a square lattice. For the \emph{discrete model},
the bond interactions take values $J_\ell = \pm 1$ at random,
and the model is characterized by the the probability $p$ of finding an
anti-ferromagnetic interaction, $J_\ell = -1$, at a given bond.
At present, the existence of a spin glass transition in these models
at $T_c \neq 0$ is still under debate.

\vskip 2mm
As first noticed by Bieche et al.~\cite{Bieche80} and further
elaborated by Nishimori~\cite{Nishimori1981,Nishimori1983},
the Ising model with random distribution of anti-ferromagnetic bonds
has a hidden $Z_2$ gauge symmetry. As discussed below, this symmetry
implies that gauge invariant observables such as the thermal energy or
the specific heat are unchanged by a certain redistribution of the
anti-ferromagnetic bonds (which may also change their number considerably).
By exploiting this invariance, Nishimori was able to obtain exact results
for special values of the parameters $p$ and
$T$~\cite{Nishimori1981,Nishimori1983}.

\vskip 2mm
The generic difficulty in simulating spin systems is that the
auto-correlation time $\tau $ increases rapidly with the physical
correlation length $\xi $ of the system, $\tau \propto \xi ^z $,
where $z$ is the dynamical critical exponent. For all local
update algorithms, $z$ is as large as $2$. This is particularly problematic at
small temperatures, when $\xi$ reaches the extension $L$ of the lattice
and the  generation of independent configurations in a Markov chain becomes
extremely cumbersome. As a consquence, the auto-correlation times must be
monitored very carefully or the algorithm might fail to be ergodic.
For a pure ferromagnet, the ground state is known explicitly: A state with
all spins parallel minimizes the energy. This knowledge of the true ground state
can be used to design an efficient algorithm which microcanonically changes
clusters of spins with the same orientation. Indeed, the so-called cluster
algorithms~\cite{swendsen87,wolff89} largely alleviate the auto-correlation
problem: the dynamical critical exponent drops to values $z \approx 0.4$
which renders practical simulations on large lattices feasible.

\vskip 2mm
So far, cluster algorithms for the frustrated Ising model do not
exist. This is mainly due to the fact that the true ground state
(and hence the structure of the physical clusters) is unknown for
a generic distribution of anti-ferromagnetic bonds. In fact, finding the
ground state is an NP hard problem in $d \ge3$. For the special case $d=2$,
Edmond's algorithm~\cite{Edmonds65a,Edmonds65b} provides a
method which computes the exact ground state in polynomial time.
This can be used to clearify the structure of the physical clusters in
special gauges, which would otherwise be obfuscated by the hidden
gauge symmetry.

\vskip 2mm
In this letter, we quantify the amount of frustration in the 2d Ising model
in a gauge invariant way by counting the fraction $\rho$ of \emph{vortices}
(non-trivial plaquettes) in a given bond distribution. First, we determine
the exact ground state energy as a function of $\rho$ using Edmond's algorithm.
In order to treat the system near criticality, we present a
novel cluster update algorithm which proposes clusters in a gauge independent
way. The specific heat as a function of the inverse temperature is explicitly
evaluated for models with different frustrations, and gauge independence
is verified.

\vskip 2mm
The partition function of the frustrated Ising model involves a summation over
all spin configurations $\{\sigma_x\}$
\be
Z \; = \; \sum _{ \{\sigma_x\} } \; \exp \Bigl\{ \sum _{\ell=\langle xy \rangle}
\beta _{\ell} \; \sigma _x \; \sigma _y \Bigr\} \; ,
\label{eq:1}
\en
where the spins located at the sites of the lattice take the values
$\sigma _x = \pm 1$. The sum in the exponent extends over all
bonds $\ell = \langle xy \rangle$ and the coupling constants $\beta _{\ell}$
are chosen positive and equal to $\beta >0$, except for a fraction $\kappa$
of the bonds where the couplings are  $(-\beta )$. In the
zero temperature limit $\beta \rightarrow \infty $, the anti-ferromagnetic
couplings induce frustration.

\vskip 2mm
It was first observed by Nishimori~\cite{Nishimori1981,Nishimori1983}
that bond distributions with vastly different values for  $\kappa $
may still share the same thermodynamical properties. This is due to a
$Z_2$ gauge symmetry, which becomes transparent if we introduce
link variables $U_\ell \; = \; \hbox{sign}(\beta _\ell)$.
With this notation, the thermal energy is given by
\be
E(T) \; = \; -\;  \Bigl\langle \sum _{\ell = \langle xy \rangle }
\; \sigma _x \, U_\ell \, \sigma _y \; \Bigr\rangle   \; .
\label{eq:12}
\en
The partition function, eq.~(\ref{eq:1}), and observables such as the thermal
energy, eq.~(\ref{eq:12}), are invariant under the following change of bonds and
spin variables:
\bea
\sigma ^\Omega (x) &=& \Omega (x) \; \sigma (x)
\nonumber \\
U ^\Omega _{\langle xy \rangle  } &=& \Omega (x) \;
U _{\langle xy \rangle  } \; \Omega^{-1} (y) \; ,
\label{eq:19}
\ena
where the \emph{gauge transformation} takes values in $Z_2$,
$\Omega (x) = \pm 1$. In order to characterize the frustration of the model,
we introduce the plaquette variable, $P(p) \; = \; \prod _{\ell\in p} U_\ell $,
constructed from the given bond background. This definition is borrowed
from lattice gauge theory, where a non-trivial value $P(p)=-1$ indicates
that a $Z_2$ vortex intersects the plaquette $p$. Notice that the variable $P(p)$
is invariant under the $Z_2$ gauge transformation (\ref{eq:19}), and the
distribution of vortices is thus the proper measure to quantify the
frustration of the model.

\vskip 2mm
With a given vortex content, there is still a large number of gauge equivalent
bond distributions which share the same physical properties.
In particle physics, the bond distribution with the minimal number of
anti-ferromagnetic bonds is known as \emph{Landau gauge},
\be
\sum _{\ell = \langle xy \rangle }
\; U ^\Omega _{\ell } \; \stackrel{\Omega }{
\longrightarrow } \; \hbox{max} \; .
\label{eq:21}
\en
For this choice of bond distribution, the ground state is always uniform,
\be
\sigma _x^{\Omega _L} \; = \; \sigma _y^{\Omega _L}  \; = \; \hbox{const.}
\; \; \forall x,y \; , \; \;  \hbox{ (Landau gauge) }  .
\label{eq:22}
\en
To see this, we can use eqs.~(\ref{eq:19}) and (\ref{eq:21}) to express
the energy of a given spin configuration  $\{\sigma \}$ in a Landau gauge
background as
\bea
E[\sigma ] &=& - \sum _{\ell = \langle xy \rangle } \sigma _x \; U_{\ell}^{\Omega_L}
\; \sigma _y =  - \sum_{\ell = \langle xy \rangle} U_\ell^{\sigma\cdot \Omega_L}
\nonumber \\
 &\ge& - \sum _{\ell = \langle xy \rangle } U ^{\Omega _L}
_{\ell} =
 - \sum _{\ell= \langle xy \rangle } \sigma _x^{\Omega _L} \;
U ^{\Omega _L} _{\ell} \;  \sigma _y^{\Omega _L} = E[\sigma^{\Omega_L}] \; ,
\nonumber
\ena
where we used the maximum condition eq.~(\ref{eq:21}) for the inequality and
the definition eq.~(\ref{eq:22}) for the uniform ground state.

\vskip 2mm
\begin{figure*}[t]
\includegraphics[height=7cm]{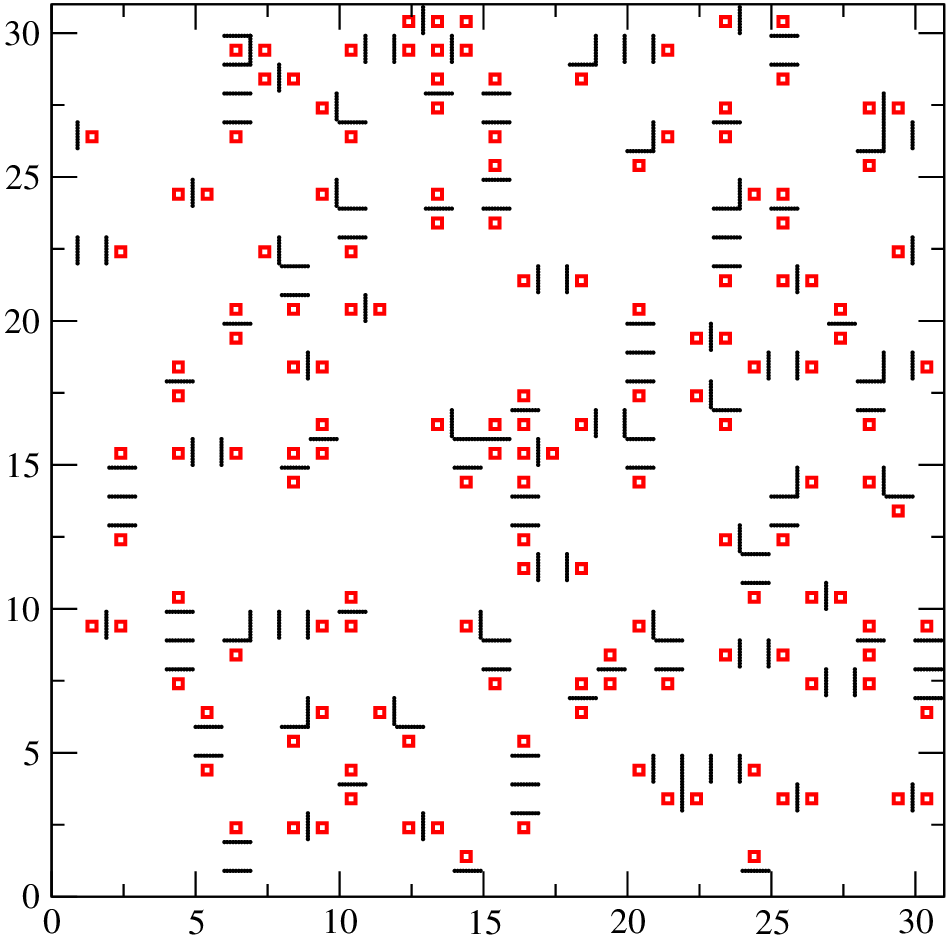} \hspace{0.5cm}
\includegraphics[height=7.5cm]{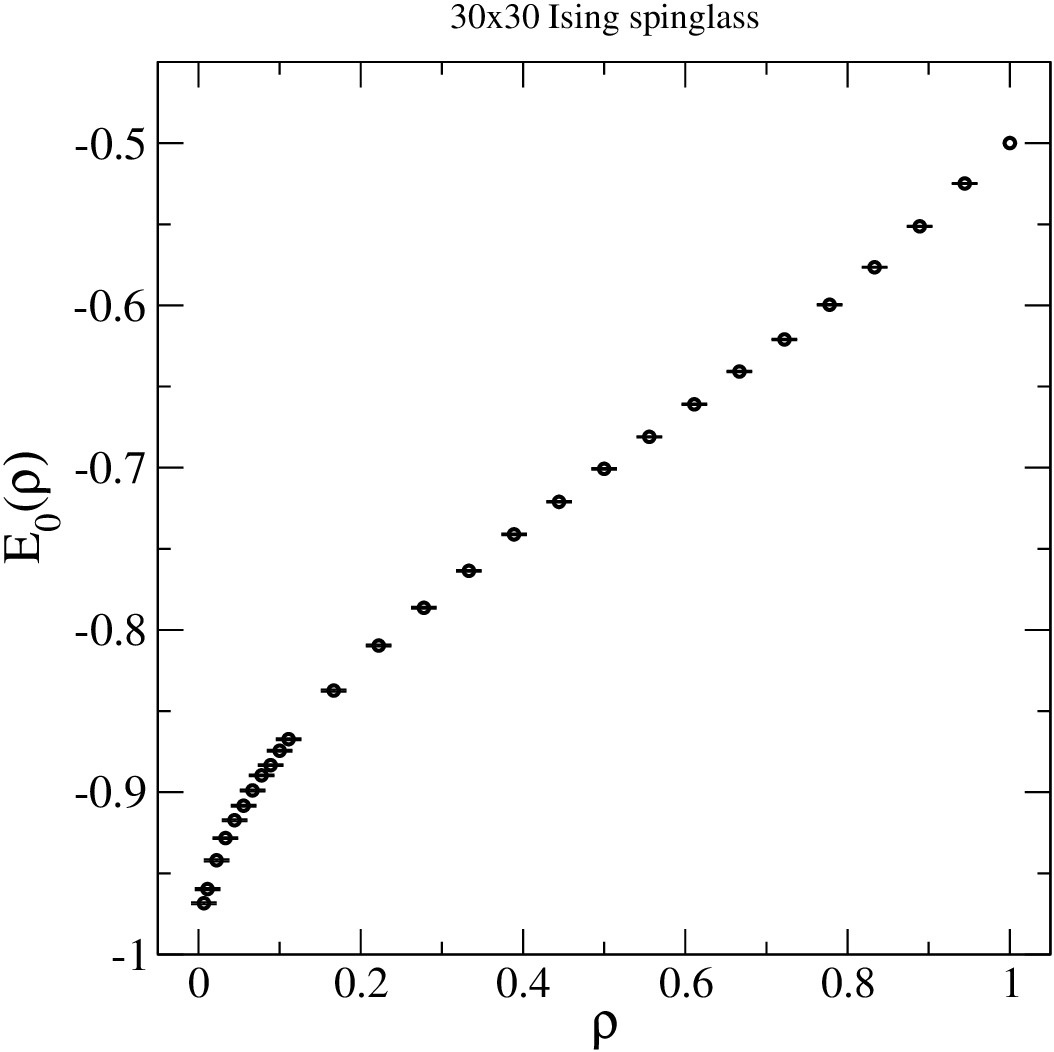}
\caption{ Left panel: $n_v=150$ randomly distributed vortices
(red squares) on a $30 \times 30$ lattice and the corresponding
minimal matching with anti-ferromagnetic bonds (black lines).
Ground state energy per link: $E_0 \; = \; -0.837778(1)$. Right panel:
ground state energy as function of the density of frustration $\rho $.
}
\label{fig:2}
\end{figure*}
Let us consider a particular distribution of anti-ferromagnetic bonds
on the lattice. The ground state energy can be obtained as follows:
(i) Calculate the position of the vortices on the lattice;
(ii) construct the minimal number $N_A$ of anti-ferromagnetic links which
are compatible with the given vortices,
(iii) obtain the gauge transformation $\Omega _x$ which casts the original
bond distribution to the one obtained in step (ii). As shown above, the
gound state in the bond distribution (ii) is uniform and its energy
can be read off directly,
\be
E _0 \; = \;  N_A \; - \; 2 \, N_\ell ,
\label{eq:23}
\en
where $ N_\ell \; = \; d \, L^d$ is the total number of bonds on the lattice.
Transforming back to the initial bond distribution, the energy $E_0$ remains
unchanged while the ground state becomes $\sigma^0_x = \Omega_x^{-1}$.
The practical difficulty in this algorithm lies in step (ii) which
is NP hard except for the case $d=2$, where Edmond's
algorithm~\cite{Edmonds65a,Edmonds65b} provides an efficient solution
in polynomial time~\cite{Bieche80}.
Figure~\ref{fig:2} shows a random distribution of $n_v=150$ vortices and the
corresponding minimal number of frustrated bonds, which were obtained with
Edmond's algorithm. Also shown is the ground state energy as function of the
vortex density $\rho  = n_v/L^2$. The data comprise an average over
100 vortex distributions for each value of $\rho $.

\vskip 2mm
In the following, we will construct a cluster algorithm which strongly
reduces auto-correlations near criticality and which, in addition,
is gauge invariant. For this purpose, we follow the derivation of the
Swendsen-Wang cluster algorithm~\cite{swendsen87}, but firstly
divide the bonds into those with ferromagnetic and anti-ferromagnetic
couplings, respectively. For the ferromagnetic bonds, we make use of
the identity
\be
\exp \{ \beta \; \sigma _x \; \sigma _y \} \; = \;  e^{ \beta } \;
\left[ \left(1-q\right) \; + \; q \; \delta _{\sigma _x
\sigma _y } \right] \; ,
\label{eq:3}
\en
with $q = 1 - e^{ -2 \beta  } $.
A similar expression holds for the anti-ferromagnetic bonds,
\be
\exp \{- \beta \; \sigma _x \sigma _y \}  =   e^{ \beta } \;
\left[ \left(1-q\right)  +  q \; (1  -  \delta _{\sigma _x
\sigma _y } ) \right]  \; .
\label{eq:4}
\en
\begin{figure*}[t]
\includegraphics[height=7cm]{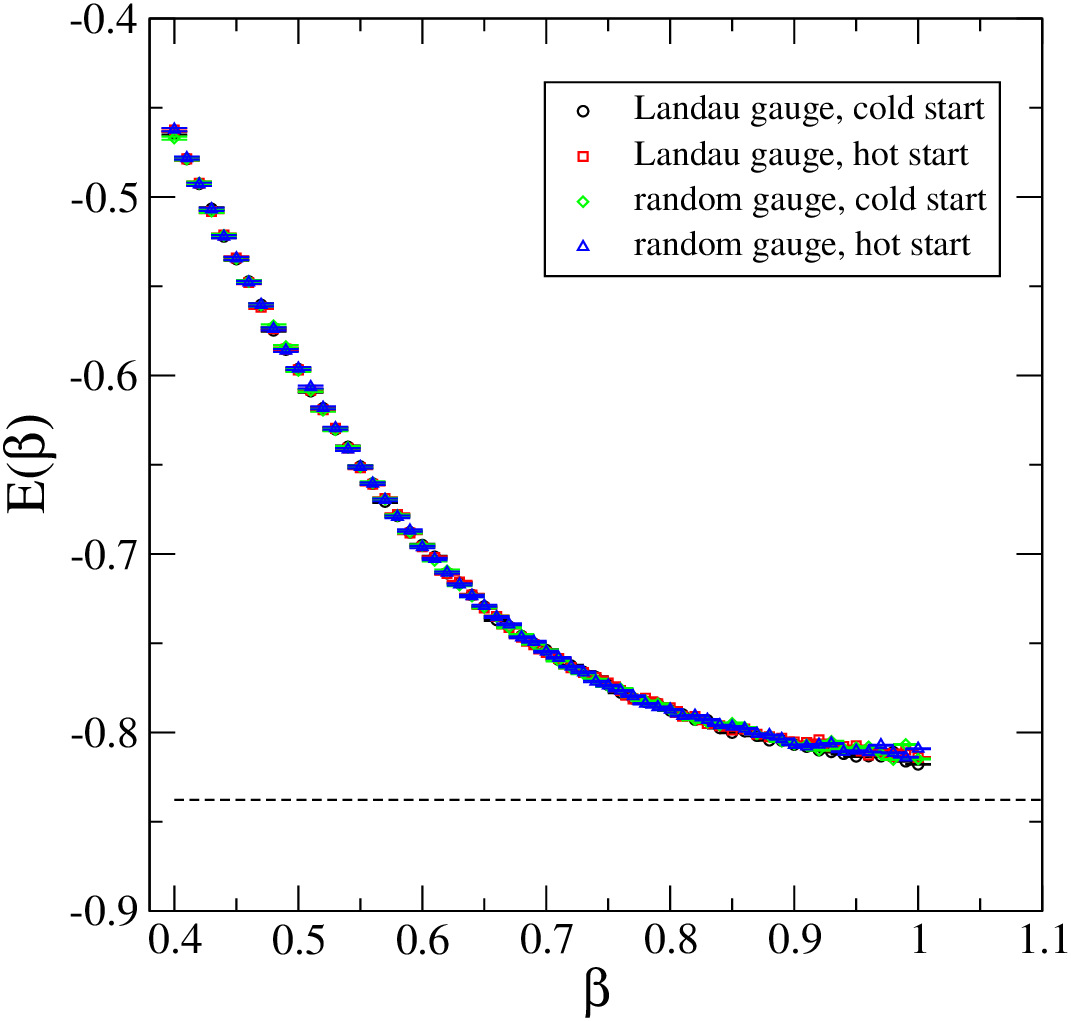} \hspace{0.5cm}
\includegraphics[height=7.5cm]{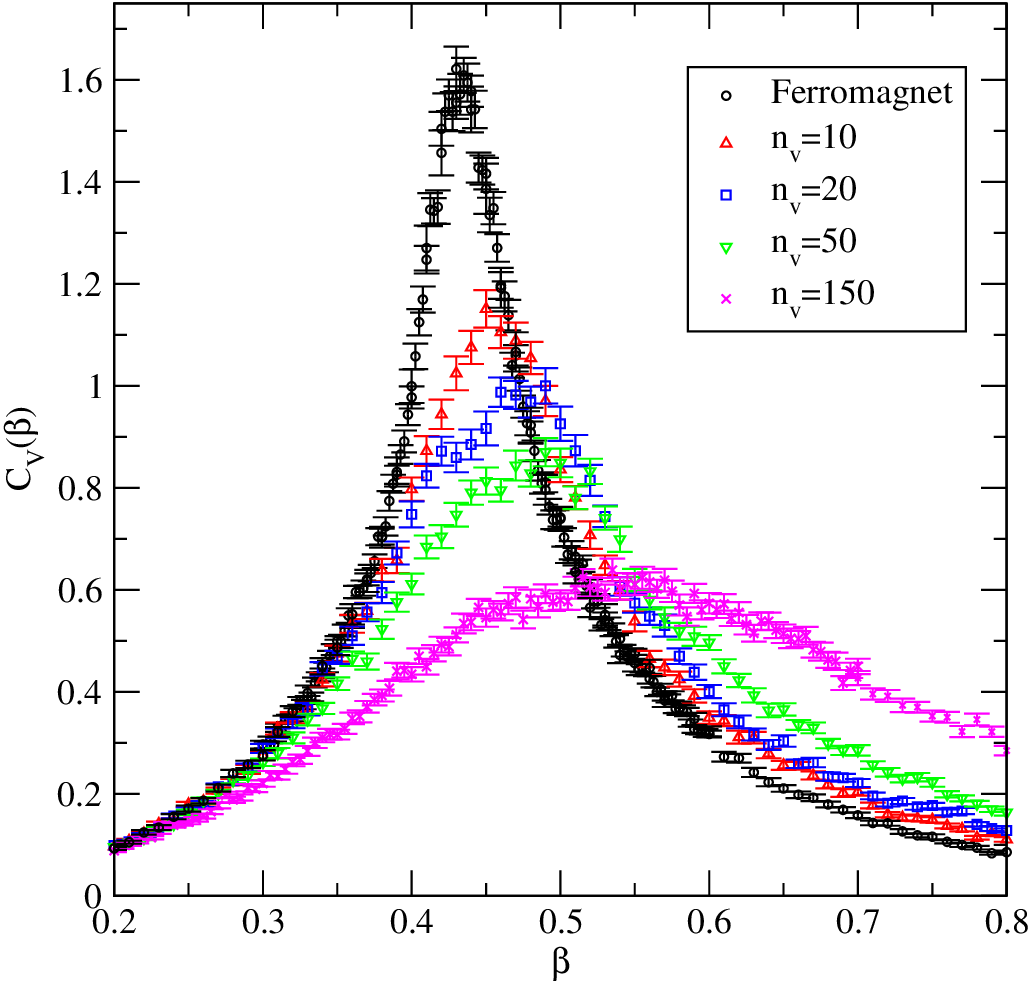}
\caption{ Left panel: Thermal energy as a function of $\beta $
for the configuration depicted in figure~\ref{fig:2}.
Right: Specific heat for a $30 \times 30$ lattice with $n_v$
randomly chosen vortices. }
\label{fig:3}
\end{figure*}
The partition function can thus be written as
\bea
Z &=&  e^{ N_\ell \, \beta } \sum _{ \{\sigma_x\} } \; \prod _\ell
\Bigl\{ \theta ( \beta _{\ell} ) \;
\bigl[ (1-q) \, + \,  q \; \delta _{\sigma _x \sigma _y } \bigr]
\label{eq:5} \\
&+&  \theta ( - \beta _{\ell} ) \;
\bigl[ \left(1-q\right) \; + \;  q \; (1 - \delta _{\sigma _x
\sigma _y } ) \bigr]  \Bigr\} \; ,
\nonumber
\ena
where $N_\ell$ is the total number of bonds on the lattice.
The bond activation variable $n \in \{0,1\}$ is introduced via the
identity
\be
a \; + \; b \; = \; \sum _{n=0}^{1} \Bigl[ a\;  \delta _{n0}  \; + \;
b \; \delta _{n1} \Bigr] \; .
\label{eq:6}
\en
The partition function is now recast to
\be
Z =  e^{ N_\ell \, \beta } \;
\sum _{ \{\sigma_x\} } \; \sum _{\{n_\ell\}} \; P( \sigma, n ) \; ,
\label{eq:7}
\en
where
\bea
 P( \sigma, n ) &=&  \prod
_{\ell=\langle xy \rangle } \biggl\{ \theta ( U_{\ell} ) \,
\Bigl[ \delta _{n_\ell \, 0} \, (1-q ) \, + \,
\delta _{n_\ell \, 1} \, q \, \delta _{\sigma _x \sigma _y } \Bigr]
\label{eq:8} \nonumber \\
&+&  \theta ( - U_{\ell} ) \,
\Bigl[ \delta _{n_\ell \, 0} \, (1-q ) \, + \,
\delta _{n_\ell \, 1} \, q \, (1  - \delta _{\sigma _x \sigma _y } ) \,
\Bigr] \, \biggr\} \; .
\nonumber
\ena
For the cluster update algorithm we perform subsequent updates of the
bond variables $n_\ell$ and the spin variables $\sigma _x$.
Let us first discuss the $n_\ell$ update and consider a specific bond
$\ell = \langle xy \rangle$. If $\ell$ is \emph{ferromagnetic},
$U_\ell = +1$, the corresponding bond variable $n_\ell$ is set to
zero if the spins attached to  $\ell$ are anti-parallel, $\sigma_x \neq \sigma_y$;
if $\sigma_x = \sigma_y$ we set $n_\ell = 1$ with probability $q$.
For the anti-ferromagnetic bonds we reverse this procedure:
If the spins attached to $\ell$ are parallel, we set $n_\ell = 0$, and
otherwise we set $n_\ell = 1$ with probability $q$. All spins connected by
activated bonds are said to be part of one cluster.
We call this algorithm a "hybrid cluster algorithm" since the spins in one
cluster generically have mixed orientations.
Notice that only parallel spins connected by a ferromagnetic bond and
anti-parallel spins connected by an anti-ferromagnetic bond can be part of
the same cluster.

\vskip 2mm
Next we turn to the spin update: In a Swendsen-Wang type~\cite{swendsen87}
update, we would randomly choose $i=\pm 1$ and assign $i$ to a randomly chosen
spin $\sigma _x$. By definition, all spins in $\sigma_x$'s cluster can
be connected to $\sigma_x$ by a continuous path of activated bonds.
There may in fact be several such paths which involve a different number
of anti-ferromagnetic bonds. However, if a particular connection of
$\sigma_x$ and $\sigma_y$ on the cluster involves an even (odd) number of
antiferromagnetic links, then \emph{every} other cluster connection of
$\sigma_x$ and $\sigma_y$ will also involve an even (odd) number.
If our target spin $\sigma_x$ is assigned the value $i$, then all cluster
spins $\sigma_y$ with even connections to $x$ must be assigned the same value
$i$, while all $\sigma_y$ with odd connections must be assigned $(-i)$,
in order to avoid a configuration with zero probabilistic weight. In practice,
we used a Wolff type variant~\cite{wolff89}: Rather than growing all clusters
on the lattice and flipping the spins with 50\% probability, we pick a
target spin, grow the corresponding cluster according to the rules above 
and the flip the entire cluster 

\vskip 2mm
The only non-trivial part in the proof of the above algorithm is
to show ergodicity, i.e.~the fact that any spin configuration can be
generated with a non-vanishing probability. In order to verify this,
it is sufficient to show that the change of a single spin occurs
with non-vanishing probability. This is possible e.g.~if the algorithm
identifies this single spin as a one-spin cluster, for which there
is always a non-zero probability.

\vskip 2mm
Let us finally demonstrate that the algorithm is indeed gauge invariant.
To this end, we assume that a particular spin $\sigma _{x_0}$ was selected
by the algorithm to be part of a cluster. A local gauge transformation,
$\Omega (x_0) = -1$, $\Omega (x\not=x_0) = 1$, changes
$\sigma _{x_0} \rightarrow - \sigma _{x_0}$. At the same time, however,
all bonds which are attached to $x_0$ also change their sign. From the cluster
rules above, it can be seen that the spin $\sigma _{x_0}$ would be part of the
same cluster even after a local gauge transformation. Every gauge
transformation can be composed as a sequence of local one-spin transformations
of the above type. Hence, the cluster growing prescription does not
depend on the gauge.

\vskip 2mm
In order to demonstrate that the algorithm yields the same results after
a gauge transformation, we have calculated the thermal energy as
function of $\beta $ for the spin-glass shown in figure~\ref{fig:2}.
In Landau gauge, there are $146$ frustrated bonds to represent the
$n_v=150$ vortices implying $\kappa _{L} = 146/1800 \approx 0.081$.
After a random gauge transformation, this
number increased to $888$ frustrated links which roughly corresponds
to a frustration density $\kappa \approx 0.5$. Indeed,
the algorithm provided the same result for these physically equivalent
systems with vastly different $\kappa$, cf.~figure~\ref{fig:3}.
We have also verified that hot and cold starts yield the same results
in both cases. In addition, the dashed line in figure~\ref{fig:3}
indicates the ground state energy of the spin glass,
which is the energy in the limit $\beta \rightarrow \infty$.
As a first application, we calculated the specific heat for a particular
spin-glass with  $n_v$ randomly distributed vortices. The result
is shown in the right panel of figure~\ref{fig:3}.
As expected, the (pseudo-)critical
point is shifted to larger values of $\beta $ if the number $n_v$ of
defects is increased. At the same time, the peak broadens.


\vskip 2mm
In conclusions, we have stressed the importance of a gauge invariant
classification of frustration. For the 2d frustrated Ising model,
the exact ground state energy was discussed as a function of the density
of gauge invariant vortices. Finally, a gauge invariant cluster
update algorithm was developed which allows for efficient computer
simulations near criticality.

\noindent {\bf Acknowledgments:}
H.R.~is supported by \emph{Deutsche Forschungsgemeinschaft} (contract DFG-Re856/4-2).


\begin{thebibliography}{sch90}

\bibitem{Binder86}
K.~Binder, A.~P.~Young,
Rev.~Mod.~Phys.~58, 801-976 (1986).

\bibitem{Alava98}
Mikko Alava and Heiko Rieger
Phys.~Rev.~E58, 4284-4287 (1998)

\bibitem{Binder79}
I.~Morgenstern and K.~Binder
 Phys.~Rev.~Lett.~43, 1615\u20131618 (1979)

\bibitem{Binder80}
I.~Morgenstern and K.~Binder
 Phys.~Rev.~B 22, 288\u2013303 (1980)

\bibitem{Young82}
N.~D.~Mackenzie and A.~P.~Young
 Phys.~Rev.~Lett.~49, 301-304 (1982)

\bibitem{Ogielski85}
Andrew T.~Ogielski and Ingo Morgenstern
 Phys.~Rev.~Lett.~54, 928-931 (1985)

\bibitem{Huse85}
David A.~Huse and I.~Morgenstern
 Phys.~Rev.~B 32, 3032-3034 (1985)

\bibitem{Rieger96}
H.~Rieger, L.~Santen, U.~Blasum, M.~Diehl, M.~J\"unger and G.~ Rinaldi,
J.~Phys.~A: Math.~Gen. 29 3939-3950 (1996)

\bibitem{Matsubara97}
F.~Matsubara, A.~Sato, O.~Koseki, T.~Shirakura
Phys.~Rev.~Lett.~78, 3237-3240 (1997)

\bibitem{Shirakura97}
T.~Shirakura, F.~Matsubara
Phys.~Rev.~Lett.~79, 2887-2890 (1997)

\bibitem{Bieche80}
I.~Bieche, R.~Maynard, R.~Rammal, J.~P.~Uhry,
J.~Phys.~A.~Math.~Gen {\bf A13} (1980)  2553.

\bibitem{Nishimori1981}
H.~Nishimori,
Prog.~Theor.~Phys.~66, 1169 (1981).

\bibitem{Nishimori1983}
Hidetoshi Nishimori and Michael J.~Stephen,
Phys.~Rev.~B 27, 5644-5652 (1983).

\bibitem{swendsen87}
  Robert~H.~Swendsen, Jian-Sheng~Wang,
Phys.~Rev.~Lett.~58, 86-88 (1987).

\bibitem{wolff89}
Ulli Wolff,
Phys.~Rev.~Lett.~62, 361-364 (1989).

\bibitem{Edmonds65a}
J.~Edmonds, Can.~J.~Math.~17 (1965) 449.

\bibitem{Edmonds65b}
J.~Edmonds, J.~Res.~NBS 69B (1965) 125.


\end{thebibliography}
\end{document}